\documentclass{PoS}

\title{Results from overlap valence quarks on a twisted mass sea}

\ShortTitle{Overlap valence on a Twisted Mass sea}

\author{N.~Garron, \\
        NIC, DESY \\
	Platanenallee 6 \\
	15738 Zeuthen, Germany \\
        E-mail: \email{Nicolas.Garron@desy.de}}

\author{\speaker{L.~Scorzato},\\
        ECT*\\
	Strada delle tabarelle, 286 \\
	38100 Trento, Italy \\
        E-mail: \email{scorzato@ect.it}}

\author{For the ETM Collaboration}

\abstract{
We present results of lattice computations using overlap fermions on a twisted mass background. $N_f=2$ full 
QCD gauge configurations have been produced by the ETM Collaboration with very light pions (down to less than
$300$ MeV), with small lattice spacing ($a \approx 0.09$ fm) and large volumes ($V/a^4=24^3\times 48$). 
By profiting of the 
good chiral properties of the overlap operator for the valence quarks, it is also possible to have a precise 
(and unquenched) determination of those physical quantities where the chiral properties are crucial. In order 
to have unquenched results, we match the valence quark mass with the sea quark mass. We also perform computations
with different quark masses in order to simulate (partially quenched) Strange and Charm quarks. 
A typical application is the computation of $B_K$, for which we present first results.
}

\FullConference{The XXV International Symposium on Lattice Field Theory\\
		 July 30-4 August 2007\\
		 Regensburg, Germany}

\begin{document}

\section{Introduction}
\label{sec:intro}
Dynamical overlap simulations are extremely expensive.
One interesting possibility 
is to use a different regularization for valence and for sea quarks.
In fact, valence quarks are much less critical from the costs point of view 
(since they only appear in the final measurements, as in the quenched case), 
but much more critical from the point of view of the symmetries.
In particular we can use the ``twisted mass'' (tm) regularization for the sea quarks,
and the ``overlap'' (ov) regularization for the valence quarks. 
This so called  \underline{\bf Mixed Actions} approach \cite{Bar:2002nr}
is very promising, since it can strongly reduce (or even
completely eliminate) the  operator mixing problem and 
has the potentiality of delivering the most precise and cost effective results
in the near future.

Violations of unitarity by lattice artifacts, which are expected,
can be studied analytically within ChPT. They may also take the form of ($O(a^2)$ suppressed) double poles,
just like in Partially Quenched QCD, but a closer inspection suggests that these might be small 
in practice \cite{Golterman:2005xa}.
Moreover, since the exact (twisted mass) sea quark matrix is available, they can also be studied numerically.
This is important in order to keep lattice artifacts under control.
A first test is reported in \cite{Bar:2006zj}.
Numerical simulations using a similar ``mixed'' approach has been reported by other collaborations also 
in this conference
\cite{Bowler:2004hs,Orginos:2007tw,Beane:2007xs,lat07:aubin,lat07:krieg,lat07:lellouch}.

In this proceedings we present our first physical results obtained with this approach.
The present analysis, which is done with limited statistics, is mainly meant to check the set-up
that we are using in order to decide on possible improvements.

The outline of this work is as follows. In the next section we describe the detailed set-up of
our computation. In section \ref{sec:pion}. we give physical results on the pion sector. In section
\ref{sec:ren}. we discuss the computation of renormalization factors, which is done using the RI-MOM method
and the Ward Identities. In section \ref{sec:bk}. we comment on our preliminary computation of $B_K$.

\section{Details of the computation}
\label{sec:details}
The gauge background that we use in the present work consists in the Twisted-Mass 
gauge configurations which have been produced by the ETM Collaboration \cite{Boucaud:2007uk,lat07:urbach}.
We summarize here the main features. We use twisted mass fermionic action at full twist with 
$N_f=2$ degenerate quarks, tree level Symanzik improved gauge action at $\beta=3.9$ which corresponds
to a lattice spacing $a\approx 0.09$ fm. The volume is $V/a^4=24^3\times 48$.
In the present study we consider a single value of the sea quark mass
$a\mu =0.004$ (the lightest available), which corresponds to a pseudo-scalar mass $m_{\pi}\approx 300$ MeV.
As mentioned in the introduction, these first results are obtained with a low statistics of 54 independent
gauge configurations. For more comments about the choice of this background for sea quarks we refer 
to \cite{Boucaud:2007uk,lat07:urbach}.

Valence quarks are described by the overlap operator \cite{Neuberger:1997fp}:
\begin{eqnarray}
D(m) &=& (\rho - \frac{am}{2}) D + m, \nonumber\\
D &=& \frac{1}{a} (1 + \frac{A}{\sqrt{A^{\dag}A}}), \qquad \qquad A = aD_W -\rho,
\end{eqnarray}
where $D_W$ is the Wilson Dirac operator and $\rho$ is a parameter that we set equal to one,
in order to optimize the locality properties of $D$ \cite{Bar:2006zj}.
Before applying the overlap operator we perform a single HYP-smearing transformation \cite{Hasenfratz:2001hp}.
The computation of the propagators is done with point-like sources chosen randomly on the whole lattice.
The inversions are performed by computing exactly the lowest 40 eigenvalues and then using
the SUMR algorithm \cite{Jagels:1994,Arnold:2003sx} with adaptive precision \cite{Chiarappa:2006hz}.
Thanks to a multiple mass procedure \cite{Jegerlehner:1996pm}, which can be extended to the SUMR solver
\cite{Chiarappa:2006hz}, we produced propagators for a wide range of bare masses down to $am=0.006$ and 
covering the Strange and Charm range. This brought a negligible loss of precision at high masses.
The cost of the computation of one full propagator is equivalent to the cost of producing a few
independent gauge configurations.
In order to understand whether the continuum limit is convenient in this approach, 
it will be important to check how the above cost ratio will
scales when $a\rightarrow 0$, at fixed physical volume.

In a previous report \cite{Bar:2006zj} we discussed a wide range of tests performed on smaller lattices and
we will not repeat them here. We only mention that the comparison of the scalar
correlator shown in \cite{Bar:2006zj} was not repeated in the larger lattice, since the low-mode
averaging \cite{DeGrand:2004qw}-- that is necessary to have a clean scalar propagator -- is rather expensive 
and we prefer to look at more physical quantities first.

\section{Results in the pion sector}
\label{sec:pion}

\begin{figure} 
\includegraphics[width=\textwidth]{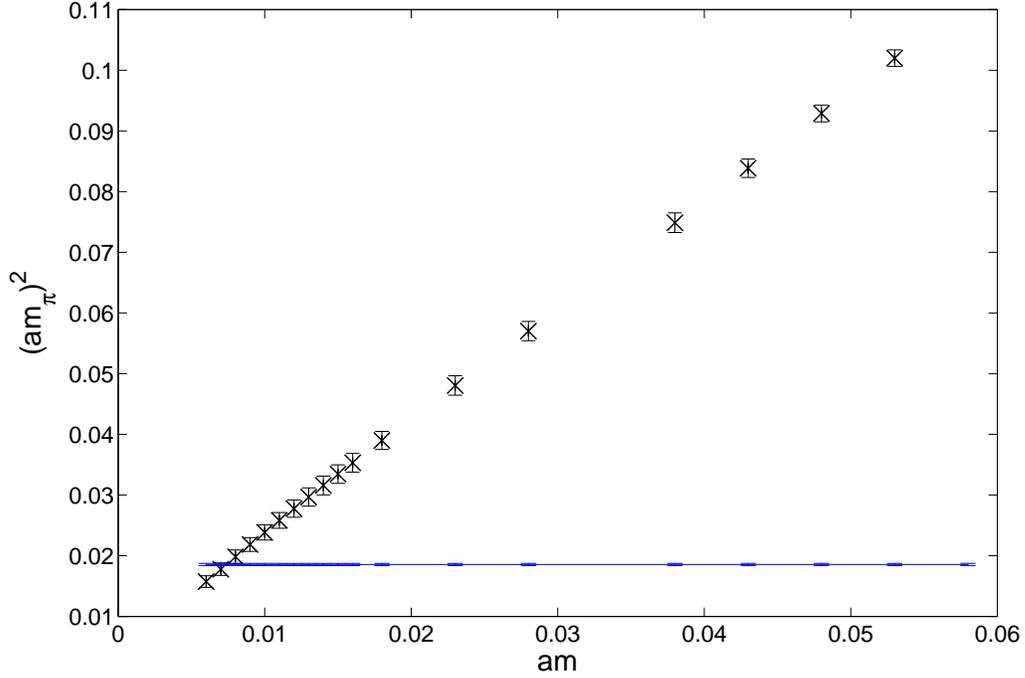} 
\caption{Matching the valence-valence and the sea-sea pion masses.} 
\label{fig:pion} 
\end{figure}

The first quantity that we consider is the pion mass, since this is also what we use to match the valence
quark mass with the sea quark mass. This is shown in Fig.~\ref{fig:pion}. 
The horizontal line (with tiny error-bars) marks the pion mass obtained in the ``unitary'' (tm-valence, tm-sea) 
set-up. From this comparison, the matching point 
is estimated to be (in the overlap bare quark mass) at $am=0.0075(10)$. The matching of one quantity
implies of course that other quantities are only matched up to lattice artifacts. The hope of this approach 
is that these are not too large in physical quantities.

The pion decay constant $f_{\pi}$ can be computed in a number of ways. The most interesting one is
the one which does not rely on any renormalization factor: 
\[
f_{\pi} = \frac{2 m}{m_{\pi}^2} |\langle 0| P | \pi \rangle |.
\]
This can be compared directly with the tm-valence tm-sea result \cite{Boucaud:2007uk}, which is also $O(a)$
improved. In this approach $f_{\pi}$ turns out to be about 10-15 \% larger than in \cite{Boucaud:2007uk}, 
at the matching point, but also the error-bars
are of the same order of magnitude, and therefore still compatible.  It is clear, from this analysis,
that some kind of noise reduction techniques as those employed in \cite{Boucaud:2007uk} would be important.

\begin{figure} 
\includegraphics[width=.5\textwidth]{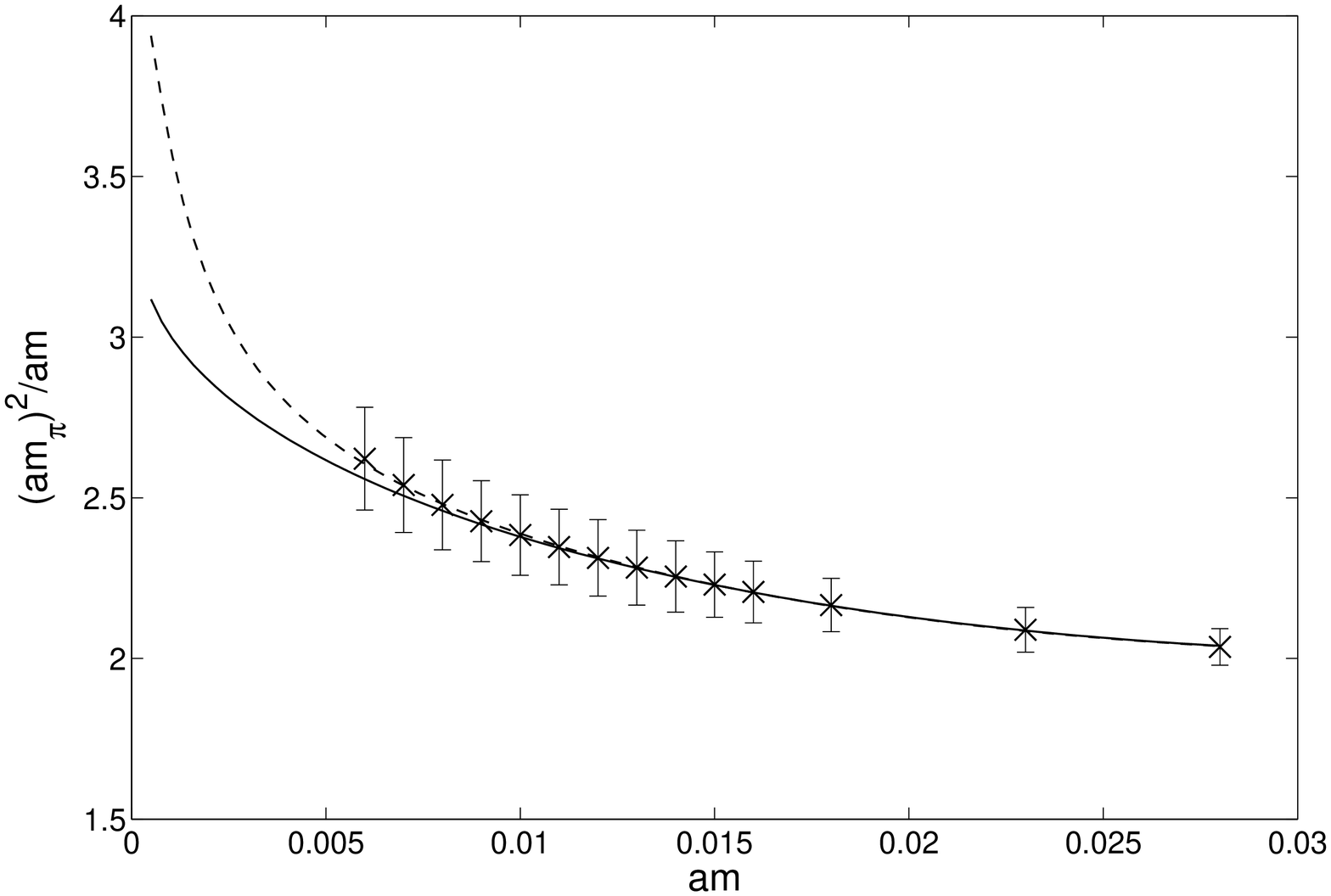} 
\includegraphics[width=.5\textwidth]{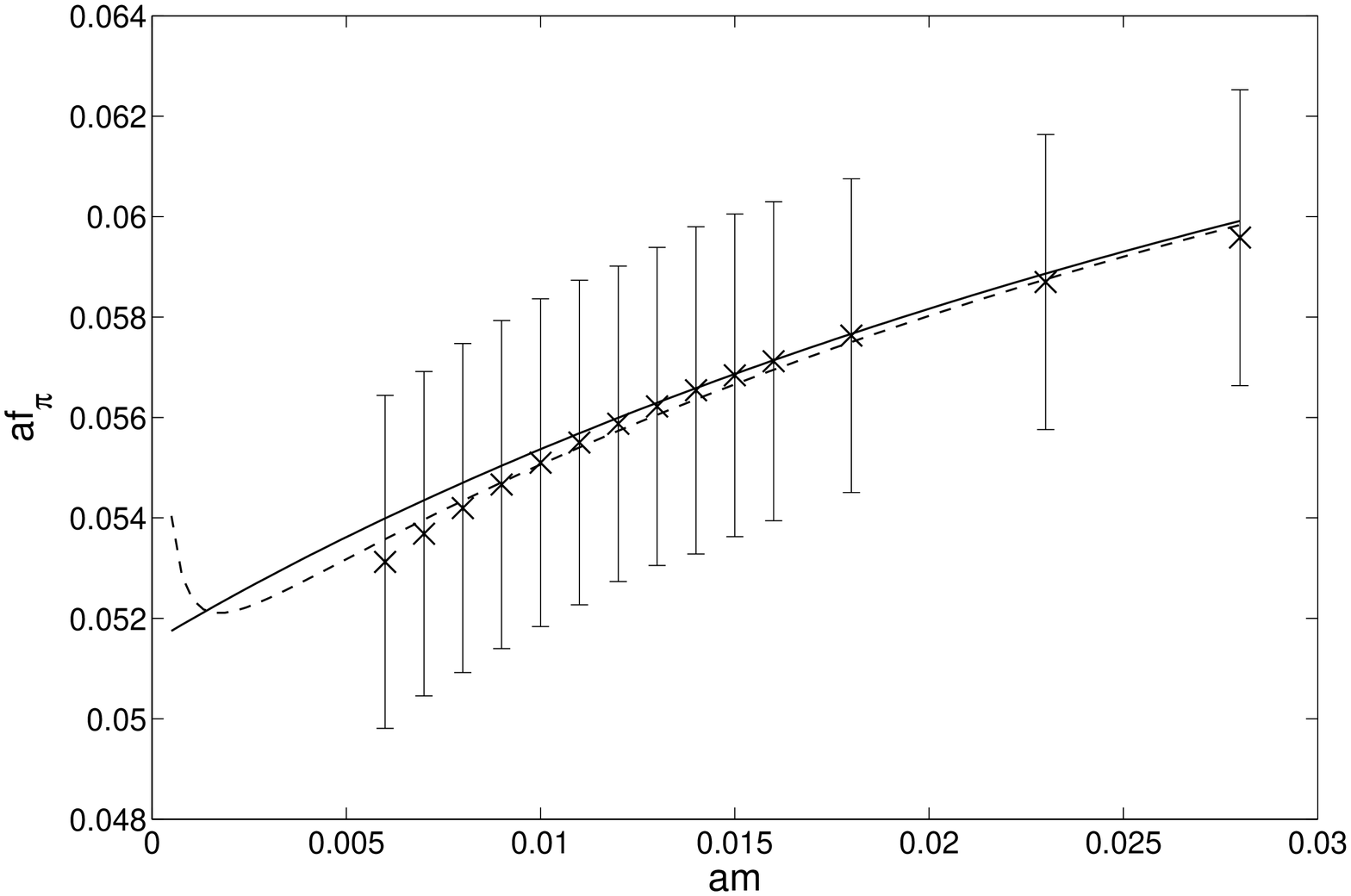} 
\caption{Fit of the data against Chiral Perturbation Theory at finite volume (dashed lines). The solid lines
are the extrapolations of the curves at infinite volume. The pion mass is plotted in a way to make the presence
of non linear corrections more evident.}
\label{fig:mfxpt} 
\end{figure}

It is also possible to compare our results for the pion masses and the pion decay constants with Chiral 
Perturbation Theory. The necessary Partially Quenched formulae have been computed in \cite{Golterman:1997st} and
the corresponding finite volume corrections in \cite{Becirevic:2003wk}.
This comparison is shown in Fig.~\ref{fig:mfxpt}. The dashed lines show the fit at finite volume, while the 
solid ones show the corresponding extrapolations at infinite volume. This gives a value of $f_0$ which is
larger than \cite{Boucaud:2007uk}, as is clear from the considerations above.

\section{Renormalization constants}
\label{sec:ren}
The renormalization factors have been computed with the RI-MOM method \cite{Martinelli:1994ty}.
This is possible since the gauge configurations had been (Landau) gauge fixed before the computation of
the propagators.

\begin{figure} 
\includegraphics[width=.5\textwidth]{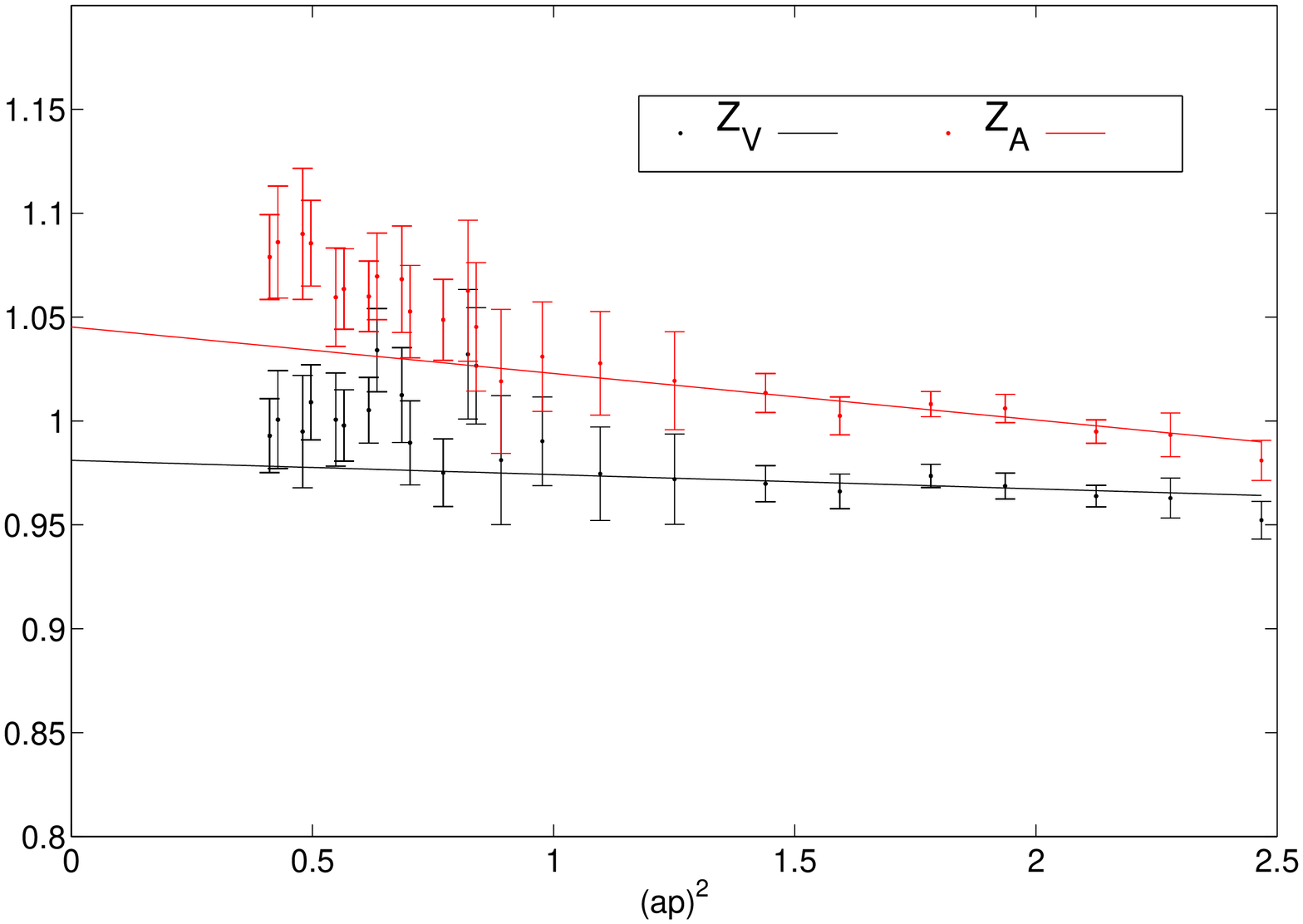} 
\includegraphics[width=.5\textwidth]{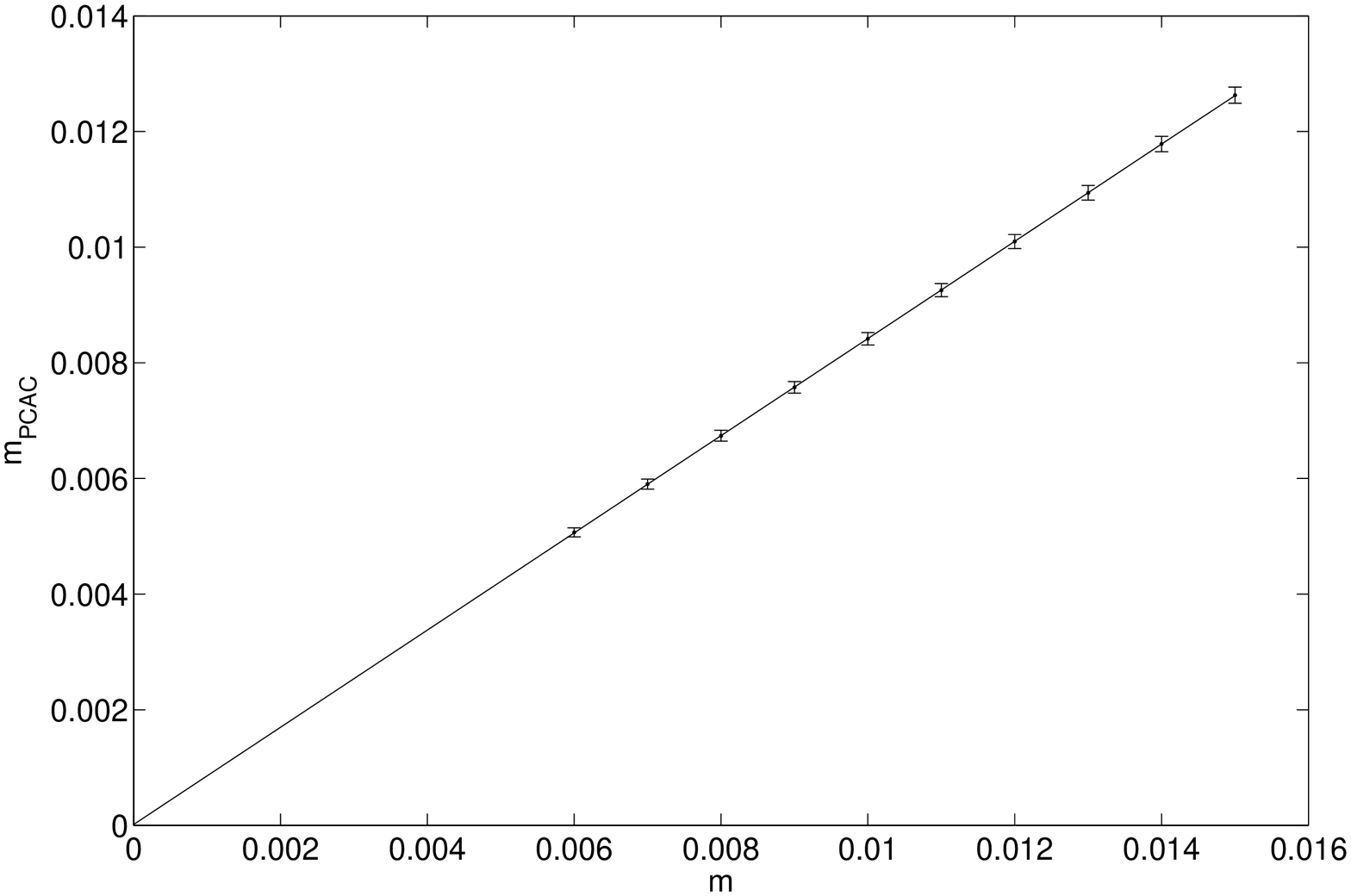} 
\caption{On the left: Plateaux for the RI-MOM determination of the renormalization factors $Z_A$ and $Z_V$.
On the right: PCAC Ward Identity.}
\label{fig:ZAZV} 
\end{figure}

It is important to note that the tree level overlap operator is different from the Wilson operator and
for $\rho=1$ the difference is significant at high momenta (which are above the cutoff, but still 
important in the RI-MOM procedure). To take this into account we define the quark field
renormalization constant $Z_{\psi}$ as:
\[
Z_{\psi}(\mu,g)=-i \frac{1}{48} 
\frac{{\rm Tr}[\; \gamma_{\nu}p_{\nu}\; S^{-1}(p)]}{
\omega(p) \sum_\mu \sin^2 ap_\mu}_{|_{p^2=\mu^2}} \nonumber \\
\qquad \qquad 
\omega(p) = \left[\sin^2(ap) + (\sin^2(\frac{ap}{2})-\rho)^2 \right]^{-\frac{1}{2}}
\]
The other definitions are unchanged with respect to \cite{Martinelli:1994ty}. 
We computed the renormalization factors for all bilinear fermionic operator and for some choices of
four fermions operators. In general we find that the chiral extrapolation is very stable, although the plateaux
are not always completely clear.
As an example, we show in 
Fig.~\ref{fig:ZAZV} (left) the plateaux for the renormalization factors of the Vector and Axial currents.
In these cases we obtain, in the chiral limit, $Z_V=0.98(5)$ and $Z_A=1.05(5)$, where the errors are
only statistical. These can be compared with the renormalization factors obtained from the PCAC Word Identities.
The relation between bare and PCAC quark masses is displayed on the right hand side of Fig.~\ref{fig:ZAZV}.
From this Ward Identity one can derive $Z_A=1.19(3)$, where the errors are only statistical.

The RI-MOM method can also be used to determine the renormalization factors of the four fermions operators. 
In particular, in the next section, we are going to use the renormalization factor of the operator 
${\cal O}_{\Delta S=2} = [\bar{s}\gamma_\nu (1-\gamma_5)d] [\bar{s}\gamma_\nu (1-\gamma_5)d]$.
The $RI$ renormalization factor can then be converted into the renormalization group invariant one 
using the anomalous dimension computed in \cite{Ciuchini:1997bw}. This gives us $Z_{B_K}^{\rm RGI}=1.48(3)$.
The momentum dependence of $Z_{B_K}^{\rm RI}(\mu)$ and $Z_{B_K}^{\rm RGI}$ in the chiral limit are 
shown in Fig.~\ref{fig:BK}.

\section{To-wards the computation of $B_K$. Comments and conclusions}
\label{sec:bk}
An obvious quantity which is particularly interesting in this approach is $B_K$, the Kaon bag parameter, which is 
related to the mixing of $\bar{K}^0$ and $K^0$ by the expression:
\[
\langle \bar{K}^0 |
{\cal O}_{\Delta S=2}(\mu)
|K^0 \rangle = 
\frac{16}{3}
M_K^2 F_K^2 B_K(\mu) 
\]
In fact a precise non perturbative determination of $B_K$ would have a strong impact on the determination
of the associated $CKM$ matrix elements. Moreover, it is only with an exactly chirally symmetric regularization 
that the operator ${\cal O}_{\Delta S=2}$ cannot mix with other operators (without need of relying on any tuning 
procedure). Finally, we have now the possibility to remove the quenching errors. 

\begin{figure} 
\includegraphics[width=.55\textwidth]{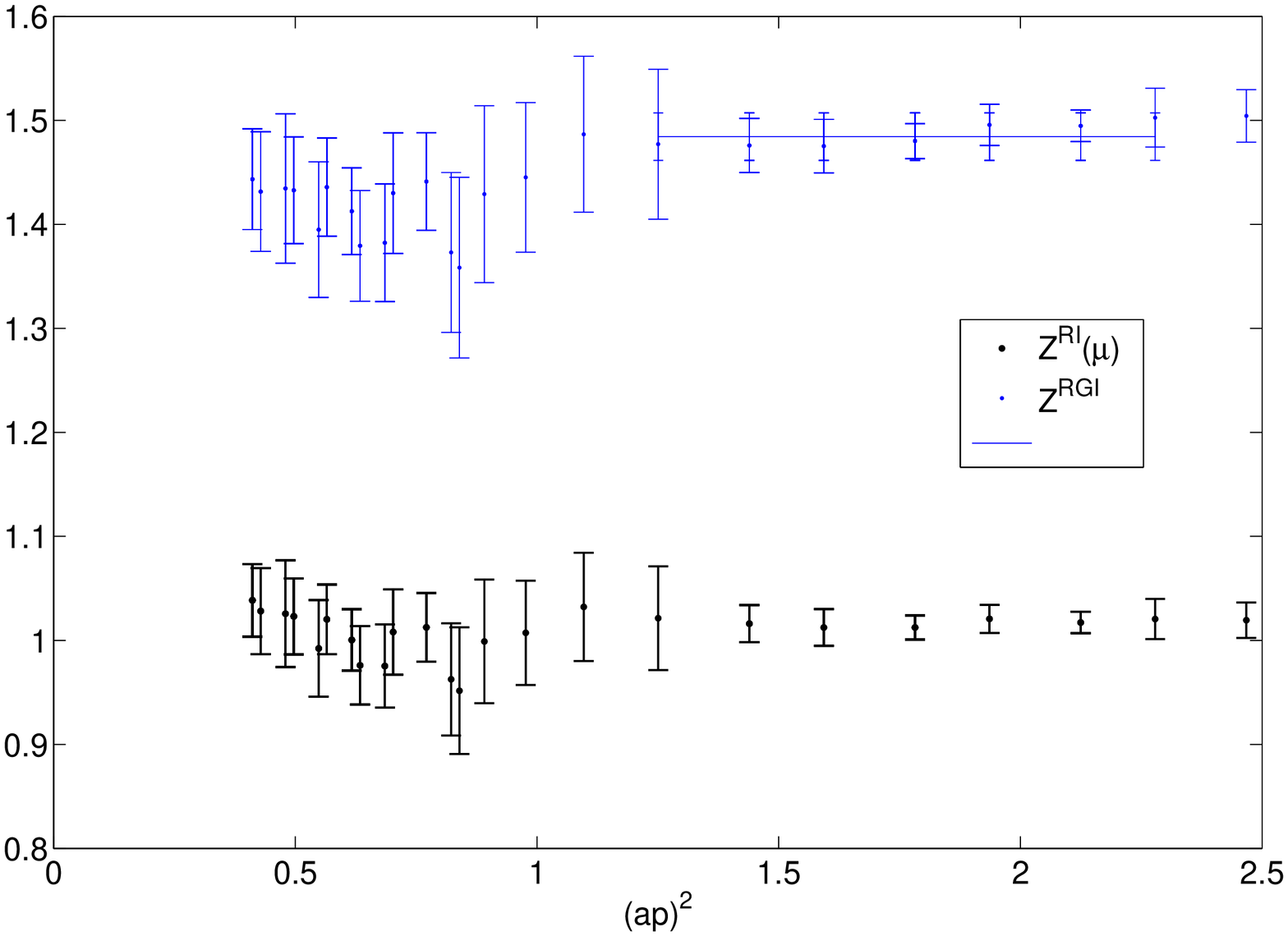} 
\includegraphics[width=.45\textwidth]{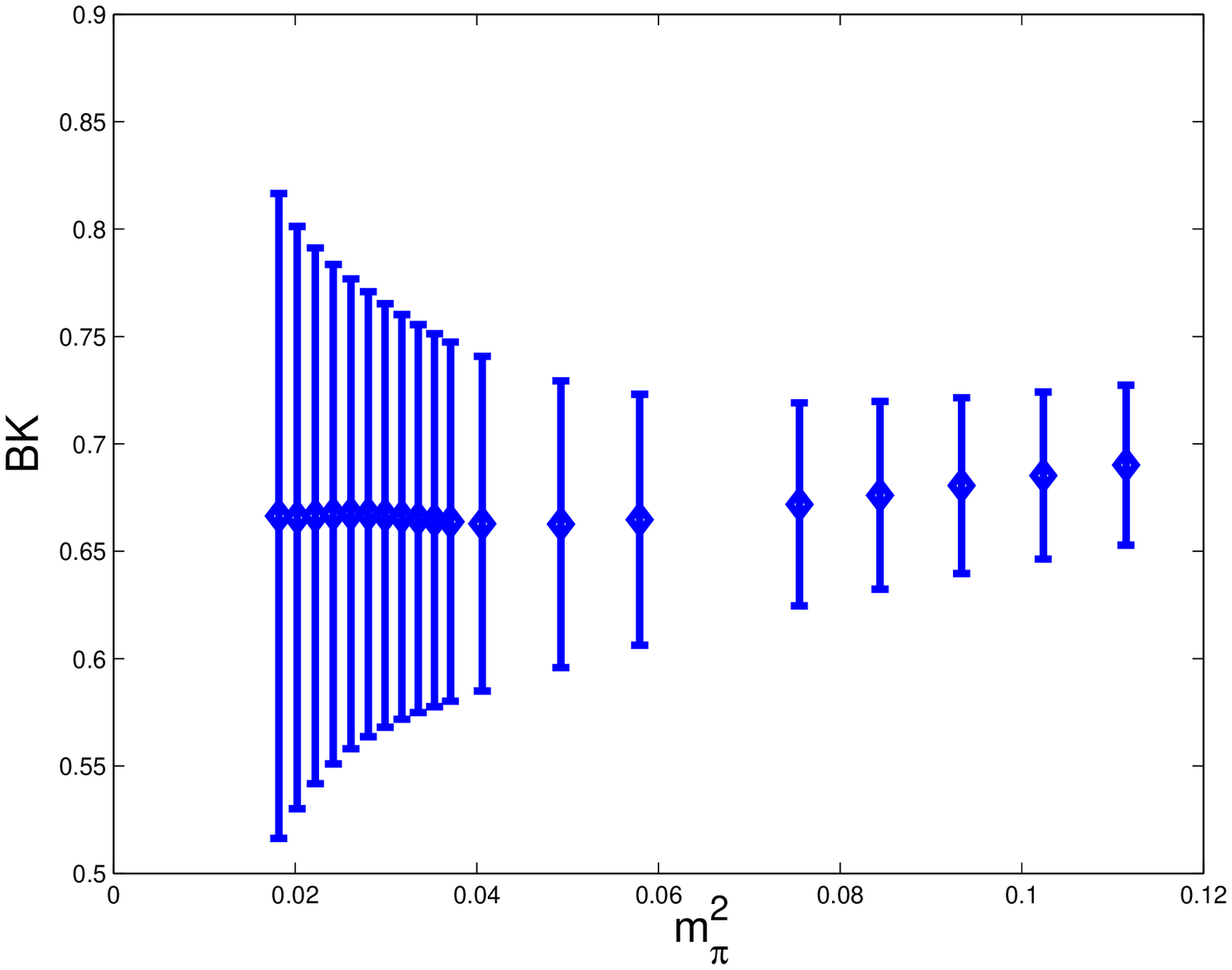} 
\caption{
Left: The momentum dependence of $Z_{B_K}^{\rm RI}(\mu)$ (bottom) and $Z_{B_K}^{\rm RGI}$ (top) 
at the chiral limit.
Right: The bare factor $B_K$ as a function of the pseudo-scalar mass. The Kaon mass corresponds here to 
$am_{\pi}^2\simeq 0.05$. 
}
\label{fig:BK} 
\end{figure}

We computed $B_K$ in a standard way employing the propagators described above. More precisely
we use the same procedure described in \cite{Garron:2003cb}, although we use lighter quark masses.
In particular it was important to use the left hand current.
Our results are shown in Fig.~\ref{fig:BK} and imply for the bare B-parameter $B_K^{\rm lat}=0.66(7)$ and 
for the renormalization group invariant one $\hat{B}_K=0.98(11)$ (errors are only statistical). 
Although the error-bars become very large at light masses, they are still reasonable at the Kaon mass,
which is relevant for $B_K$. Nevertheless, some kind of noise reduction technique would be probably helpful
and we are currently exploring those used in \cite{Boucaud:2007uk}.

Comparison with ChPT has been performed using the formulae in \cite{Golterman:1997st,Becirevic:2003wk},
and the inclusion of appropriate lattice artifacts can be done following the procedure in  \cite{Chen:2007ug}.

\acknowledgments We wish to thank the members of the ETM Collaboration who have contributed to the
results presented here and in particular O.~B\"ar, K.~Jansen, S.~Schaefer, A.~Shindler.
L.S. acknowledges INFN for support and NIC/DESY for hospitality.
Numerical work has been done in the SGI altix at HLRB (M\"unchen), in the IBM p690 at ZIB (Berlin)
and in the BEN cluster at ECT* (Trento). We are grateful to A.~Vladikas and V.~Lubicz for useful 
discussions about the RI-MOM method.

\end{document}